# Angular Characteristics of Attenuation and Temporal Dispersion Introduced by Single Deciduous Bush in Winter Conditions at FR2 Band

Tomasz Grasza
*Military University of Technology*
*Faculty of Electronics,*
Warsaw, Poland
tomasz.graszka@wat.edu.pl

Jarosław Wojtuń
*Military University of Technology*
*Faculty of Electronics,*
Warsaw, Poland
jaroslaw.wojtun@wat.edu.pl

Cezary Ziółkowski
*Military University of Technology*
*Faculty of Electronics,*
Warsaw, Poland
cezary.ziolkowski@wat.edu.pl

Jan M. Kelner
*Military University of Technology*
*Faculty of Electronics,*
Warsaw, Poland
jan.kelner@wat.edu.pl

Rajeev Shukla
*National Institute of Technology,*
*ECE Department*
Durgapur, WB, India
rksingh@ietlucknow.ac.in

Aniruddha Chandra,
*National Institute of Technology,*
*ECE Department*
Durgapur, WB, India
achandra.ece@nitdgp.ac.in

Radek Zavorka
*Brno University of Technology,*
Brno, Czechia
xzavor03@vutbr.cz

Jiri Blumenstein
*Brno University of Technology,*
Brno, Czechia
blumenstein@feec.vutbr.cz

Ales Prokes
*Brno University of Technology,*
Brno, Czechia
prokes@vutbr.cz

***Abstract*—Vegetation-induced attenuation and temporal dispersion remain important impairments in millimeter-wave radio propagation, particularly in the frequency range 2 (FR2) band, where highly directional links are employed. Existing vegetation attenuation models, including ITU-R P.833, predominantly address scenarios involving trees or vegetation clusters and are typically derived for the main radiation direction of the transmitting antenna. This paper presents an experimental study of angular-dependent attenuation and temporal dispersion introduced by a single deciduous bush under winter conditions. Measurements were conducted on a university campus at 28 GHz and 38 GHz using a wideband channel-sounding testbed operating from 1 MHz to 44 GHz. The bush, approximately 3 m in height, was located at the center of a circular measurement trajectory with a radius of 10 m, along which the receiver was positioned at angular intervals of approximately 8 deg. Directional antennas with 7–9 deg half-power beamwidth were employed, and reception angles from 0 to 200 deg relative to the transmitter-bush axis were investigated. The results demonstrate significant angular variability in both excess attenuation and temporal dispersion, indicating substantial contributions from scattering outside the main propagation direction. Differences between the two carrier frequencies are analyzed, highlighting frequency-dependent attenuation mechanisms even in leaf-off conditions. The presented findings extend existing single-vegetation models by incorporating angular characteristics and temporal effects, providing valuable insights for the design and modeling of FR2 wireless links in sparse-vegetation environments.**

This work was supported in part by the National Science Centre, Poland, under the OPUS call in the Weave program under research project no. 2021/43/I/ST7/03294 acronym 'MubaMilWave', in part by the Czech Science Foundation under project no. 23-04304L; in part by the Military University of Technology under research project no. UGB/22-128/2026/WAT on "Analysis of the possibility of using modern technologies in communication systems and electronic warfare", and in part by the MeitY, Government of India, through the Chips-to-Startup program under project no. EE-9/2/2021-R&D-E.

## I. Introduction

Millimeter-wave (mmWave) frequency bands play a crucial role in contemporary and future wireless communication systems. The high attenuation and short range of radio systems operating in the frequency range 2 (FR2) enable multiple uses of the same radio resources in close proximity. On the other hand, the FR2 bands allow for the allocation of large contiguous bandwidths, enabling very high data rates. At the same time, radio propagation at these frequencies is highly susceptible to environmental obstructions, including vegetation, which may introduce significant excess attenuation and temporal dispersion. Accurate characterization of vegetation-induced effects is therefore essential for reliable link budget estimation and channel modeling [1], [2].

Most existing vegetation attenuation models, such as the one presented in the recommendation ITU-R P.833-10 [3] of the International Telecommunication Union (ITU), are primarily derived for scenarios involving trees or dense vegetation clusters and are typically limited to the main radiation direction of the transmitting antenna. However, in many practical deployment scenarios—such as campus, suburban, or urban green areas—wireless links often interact with isolated vegetation objects, including individual bushes or small trees. In such cases, angular-dependent scattering, diffraction, and non-line-of-sight (NLOS) propagation mechanisms may contribute significantly to the received signal, especially when highly directional antennas are used [4], [5], [6], [7].

This paper presents an experimental investigation of angular-dependent attenuation and temporal dispersion introduced by a single deciduous bush under winter (leaf-off) conditions. The study focuses on two carrier frequencies, 28 GHz and 38 GHz, which are representative of current and future mmWave communication systems. By conducting measurements along a circular trajectory around the vegetation object, the angular variability of the propagation channel is analyzed beyond the main propagation direction.

This research was funded in part by the National Science Center (NCN), Poland, grant no. 2021/43/I/ST7/03294 (MubaMilWave).

The objective of this work is to extend existing single-vegetation propagation models by incorporating angular characteristics and temporal effects. Thereby, providing new insights into frequency-dependent attenuation mechanisms in sparse-vegetation environments. Such results are particularly relevant for the design and modeling of directional mmWave links operating in realistic outdoor scenarios [8], [9], [10], [11].

In our previous studies, we investigated angular propagation effects caused by a single tree at 80 GHz in an open-area environment across different seasons [5], [10]. In that case, the angular resolution of the measurements was relatively coarse, which may have introduced significant uncertainty in the derived attenuation models based on limited directional observations [5], [11]. In the present measurement campaign, we focus on lower mmWave frequency bands (i.e., 28 GHz and 38 GHz). On the other hand, we substantially improved the angular resolution of the measurement procedure. The results presented in this paper correspond to winter (leaf-off) conditions. Future work will extend the study to other seasons to quantify seasonal variability and further refine angular-dependent vegetation-propagation models.

The remainder of the paper is organized as follows. Section II describes the measurement scenario and the experimental testbed, including the measurement geometry and equipment configuration. Section III introduces the considered channel metrics and their definitions, with particular emphasis on mean delay and root-mean-square (RMS) delay spread. Section IV presents and discusses the measurement results, including angular characteristics of mean delay, RMS delay spread, and path loss (PL) at 28 GHz and 38 GHz. Finally, Section V concludes the paper and outlines directions for future work.

## II. Measurement Scenario and Testbed

The measurement campaign was conducted in an outdoor environment on the campus of the Military University of Technology (WAT) in Warsaw, Poland. The objective of the measurements was to characterize angular-dependent attenuation and propagation properties of millimeter-wave signals in the presence of a single vegetation object. The general measurement concept and the angular observation geometry are illustrated in Fig. 1 and Fig.2.

The investigated object was a deciduous bush with a height of approximately 3 m and a dense branch structure, fully devoid of leaves, which corresponds to typical winter conditions. Measurements were performed under stable atmospheric conditions. During the campaign, the air temperature was −9°C (with a wind chill of −15°C), no precipitation was observed, and the measurement area was uniformly covered with a snow layer of approximately 10 cm. The relative humidity was 81%, the atmospheric pressure was 1037 hPa, and the wind speed was approximately 14 km/h. These conditions limited short-term channel fluctuations and reduced the influence of environmental dynamics on the measurement results.

The bush was located at the center of a circular measurement trajectory with a radius of 10 m, measured from the outer boundary of the irregularly shaped bush. The transmitting antenna (TX) was placed at a fixed reference position corresponding to 0°, while the receiving antenna (RX) was moved along the circular path around the bush. Measurements were performed with an angular step of 8°, resulting in a total of 25 RX positions and covering an angular range from 0° to 200° relative to the TX–bush axis. Both antennas were mounted at a height of 1.6 m above ground level and operated without mechanical down-tilt.

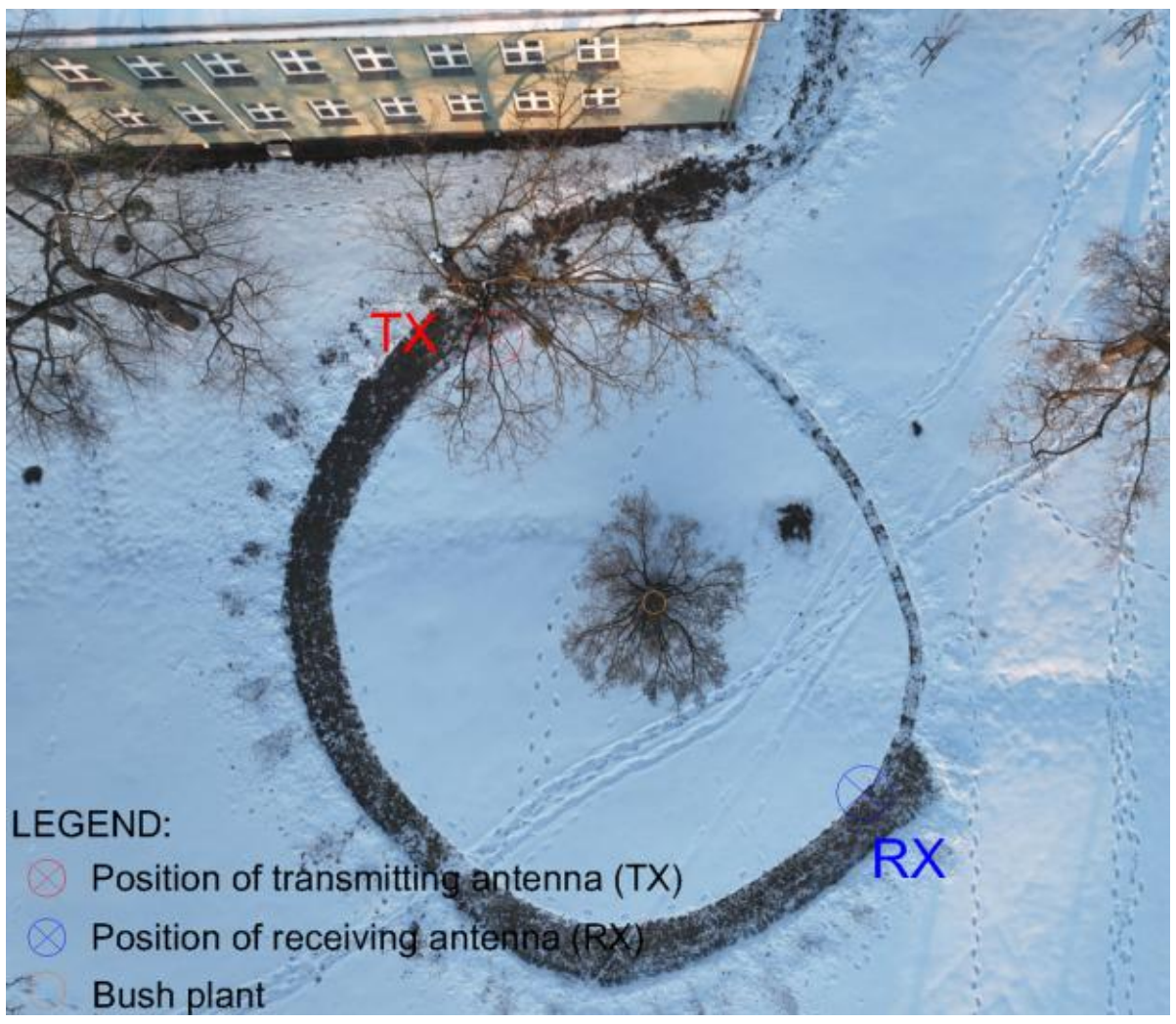

Fig. 1. Measurement scenario illustrating positions of TX and RX antennas along circular trajectory around investigated object – deciduous bush in winter conditions. Measurement points, antenna locations, and observation angles are indicated.

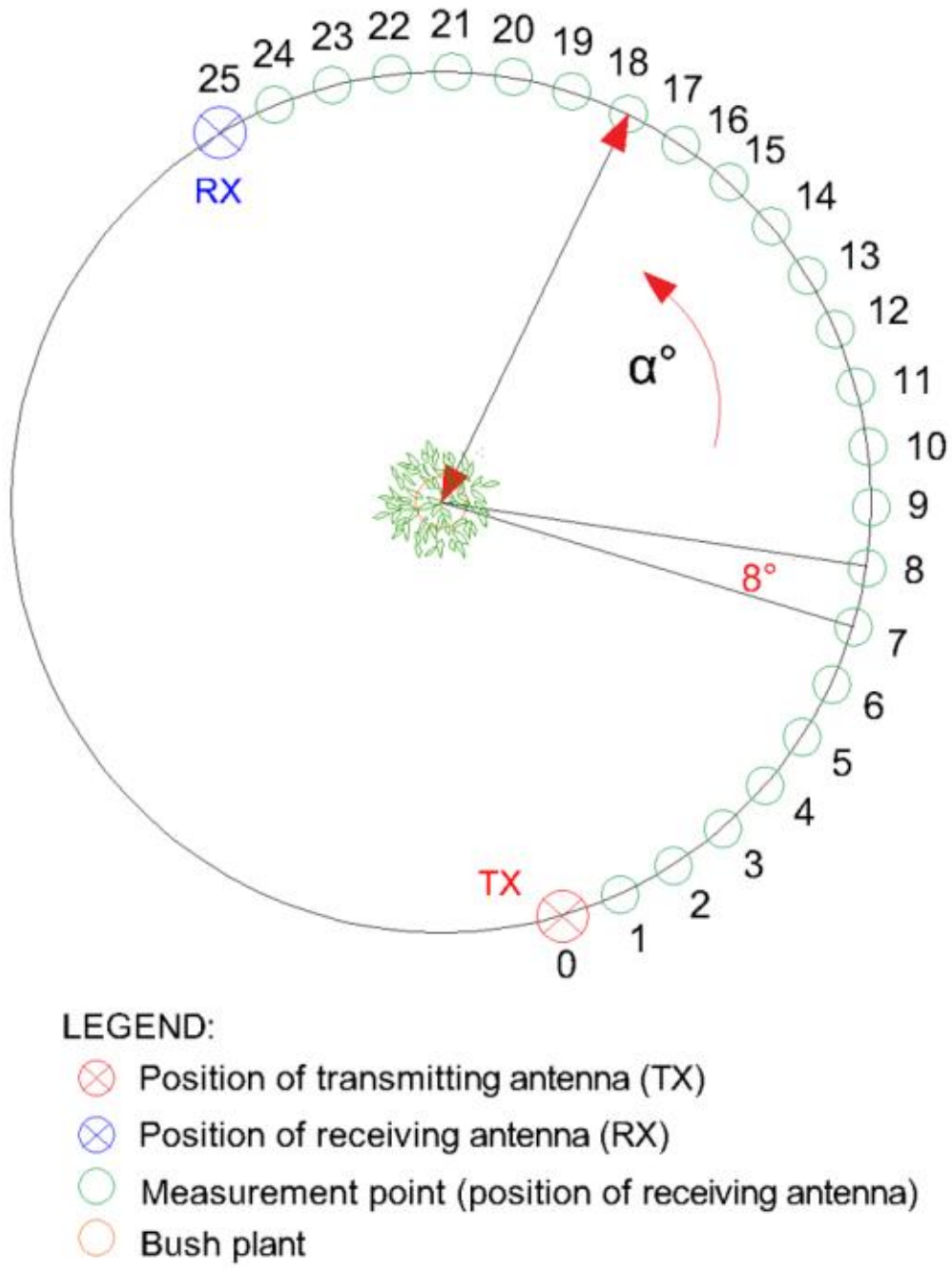

Fig. 2. Measurement scenario illustrating positions of TX and RX antennas along circular trajectory around investigated object.

Directional horn antennas with a 3 dB half-power beamwidth (HPBW) of 7° in the E-plane and 9° in the H-plane were used at both the transmitting and receiving sides [12]. This antenna configuration enabled investigation of propagation effects outside the main radiation lobe, including scattering and diffraction from the bush structure. Measurements were conducted at two carrier frequencies, namely 28 GHz and 38 GHz.

On the transmitting side, the chain included a Ka-band power amplifier operating in the 20–44 GHz frequency range,

with a gain of 45 dB (model SBP-2034434526-2F2F-S1) [13]. The connection between the vector signal generator and the power amplifier was realized using a 6 m coaxial feeder cable. Subsequently, the power amplifier was connected to the transmitting antenna via a 1 m coaxial feeder cable. On the receiving side, the antenna was connected directly to the signal analyzer via a coaxial feeder cable, with no intermediate components.

For each angular position of the RX, the channel response was recorded over a duration of 10 ms, corresponding to 1000 sounding snapshots. This enabled temporal averaging as well as the assessment of short-term channel variability. The detailed measurement setup and antenna configuration are shown in Fig. 3.

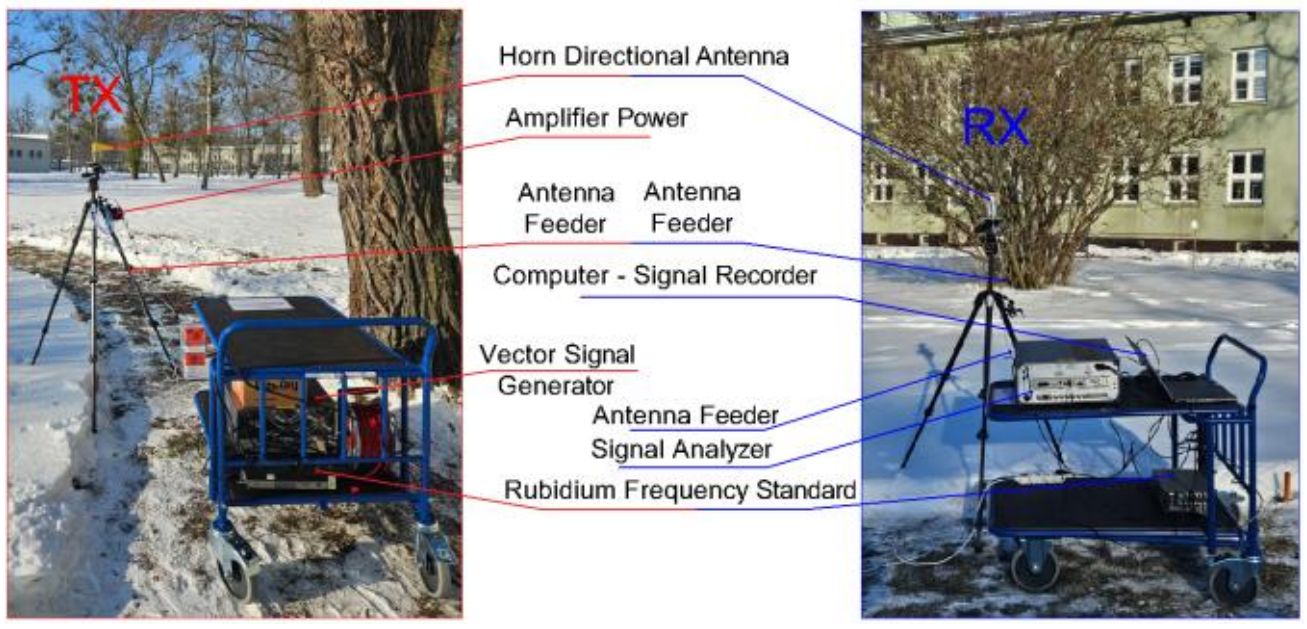

Fig. 3. TX and RX parts of measurement setup.

A 900 MHz bandwidth frequency-modulated continuous-wave (FMCW) waveform was transmitted, and complex in-phase/quadrature (IQ) samples were recorded at a sampling rate of 1.2 GHz. The signal generator output power was set to −20 dBm for both carrier frequencies, ensuring comparable measurement conditions. The adopted measurement scenario enabled a detailed analysis of angular-dependent attenuation and temporal dispersion effects introduced by a single vegetation object under winter conditions.

The simulated radiation patterns of the horn antenna at 26.5 GHz in E- and H-planes are shown in Fig. 4 [12]. Based on the obtained patterns, the effective angular coverage of the antenna was determined to be approximately 20°. This angular range was defined as the region around the boresight direction in which the antenna gain remains within 10 dB of its maximum value. Although the 3 dB beamwidth equals 7° in the E-plane and 9° in the H-plane, the wider 10 dB criterion was adopted in order to account for significant off-boresight signal contributions observed in the subsequent angular-dependent propagation measurements.

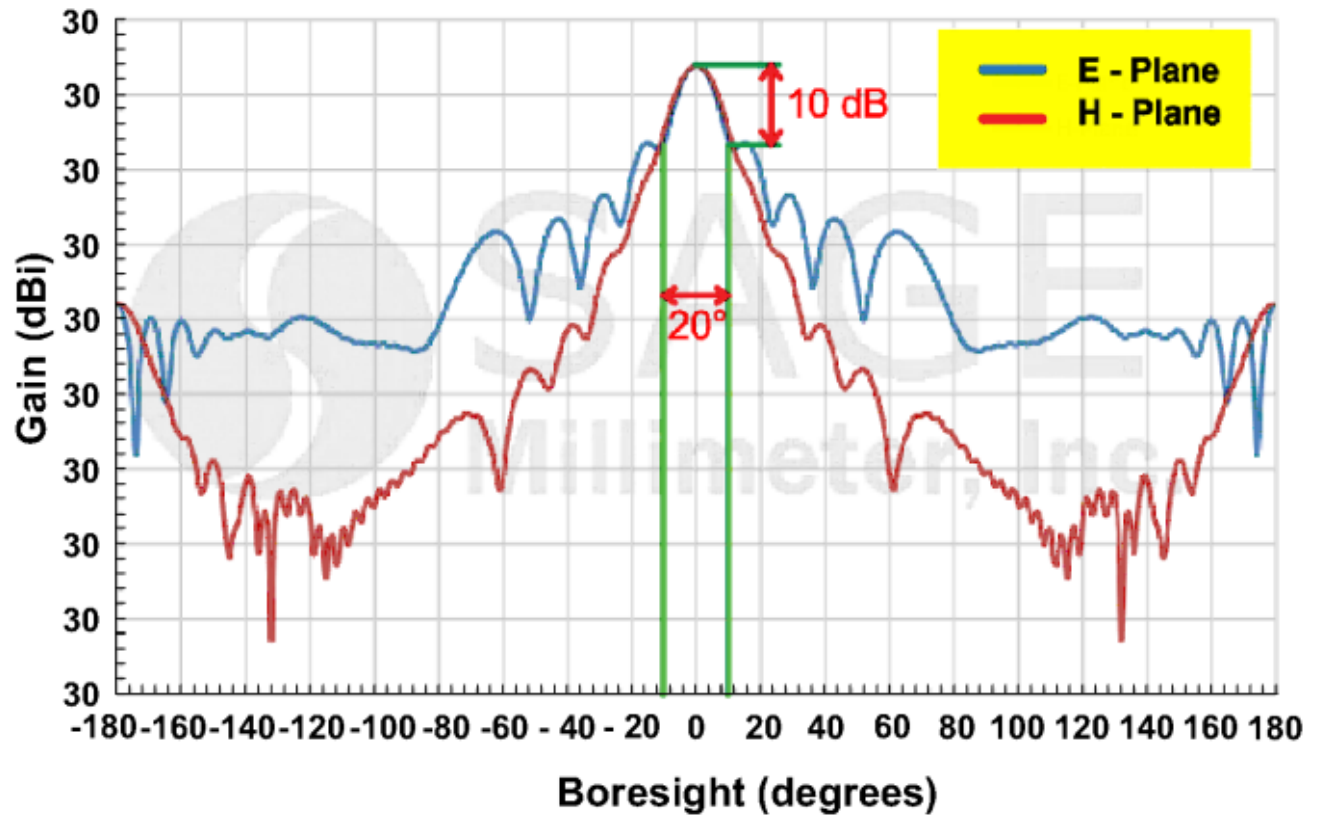

Fig. 4. E- and H-plane radiation patterns of the horn antenna at 26.5 GHz.

Fig. 3 depicts the angular-domain measurement scenario, in which the transmitting antenna remains fixed and is directed toward the investigated bush, while the receiving antenna moves along a circular trajectory around the object and is oriented toward the bush at each measurement position.

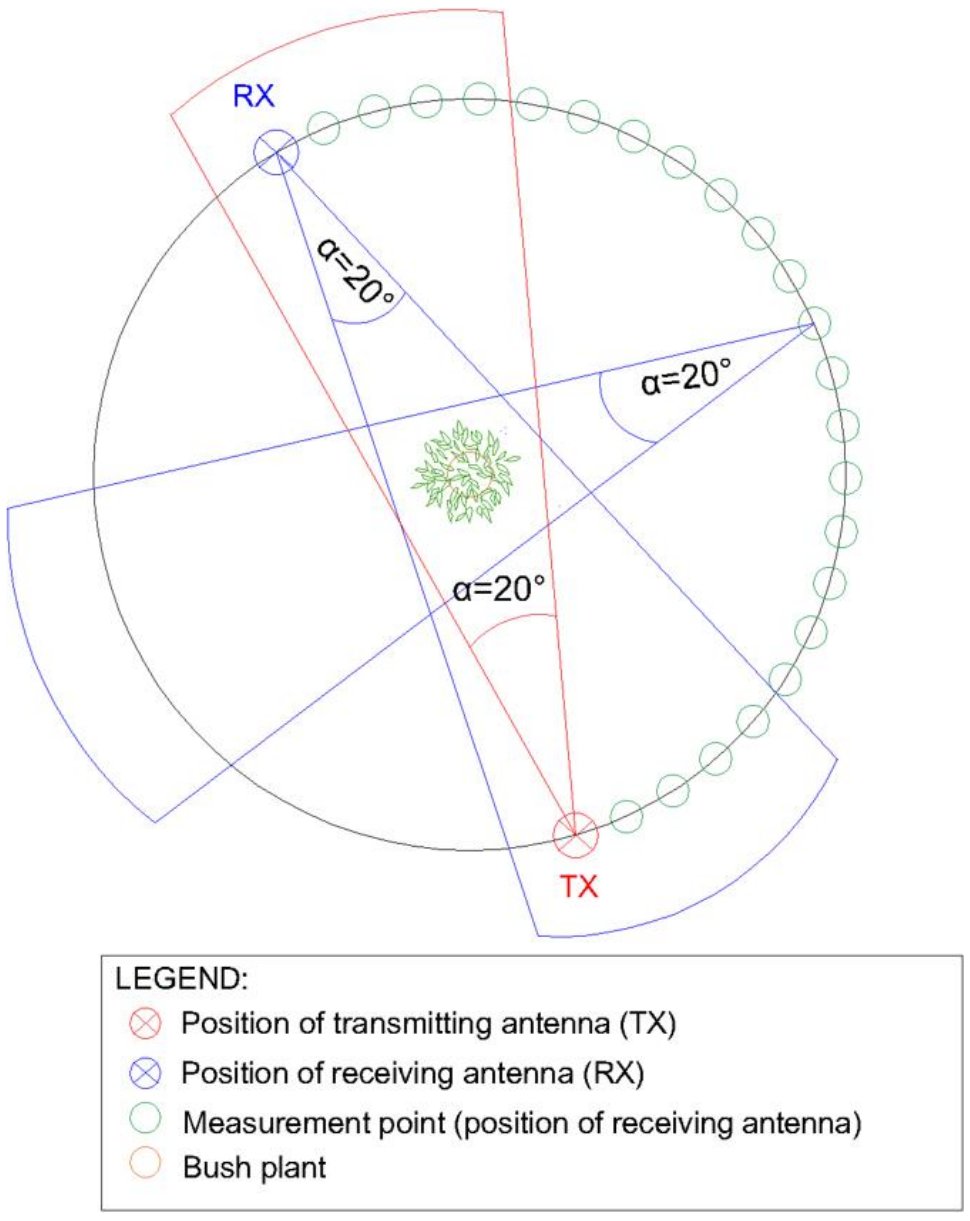

Fig. 5. Angular coverage of TX and RX antenna beams in measurement scenario for selected measurement points.

The angular positions corresponding to the strongest overlap of the antenna beams define the reference propagation direction and serve as a baseline for interpreting delay-domain channel metrics. As the angular offset from this direction increases, the contribution of the direct component gradually decreases, while scattered and diffracted components associated with interactions with the bush structure become increasingly significant.

## III. Channel Metrics and Definitions

To quantitatively characterize the temporal properties of the millimeter-wave propagation channel in the presence of a single vegetation object, selected delay-domain metrics were extracted from the measured wideband channel impulse responses. In addition to the temporal analysis, the PL was evaluated as a function of the TX-bush-RX angle, $\alpha$, to assess the attenuation introduced by the vegetation object. In this study, the analysis focuses on the mean excess delay and the RMS delay spread, which provide a compact description of the channel's temporal structure and the degree of multipath dispersion.

### *A. Mean Delay*

The mean excess delay represents the first moment of the power delay profile (PDP) and characterizes the average arrival time of the received multipath components relative to the first detectable path. It is defined as [11], [14]

$$\bar{\tau} = \frac{\sum_i \tau_i P(\tau_i)}{\sum_i P(\tau_i)} \tag{1}$$

where $\tau_i$ and $P(\tau_i)$ denote the delay and power of the $i$th multipath component, respectively. This metric is useful for assessing the contribution of delayed signal components arising from scattering by the investigated vegetation object.

## B. RMS Delay Spread

The RMS delay spread quantifies the temporal dispersion of the channel by measuring the spread of multipath components around the mean excess delay. It is given by [11], [14]

$$\sigma_\tau = \sqrt{\frac{\sum_i (\tau_i - \bar{\tau})^2 P(\tau_i)}{\sum_i P(\tau_i)}} \quad (2)$$

The RMS delay spread is a key parameter for evaluating the impact of multipath propagation on wideband communication systems, particularly in millimeter-wave links employing highly directional antennas [10], [11], [14].

# IV. Results and Discussion

## A. Mean Delay

Fig. 6 shows the mean excess delay as a function of the observation angle. The lowest delay values occur in the region of the strongest alignment between the TX and RX antenna directions, confirming the dominance of the quasi-line-of-sight (QLOS) component. For larger angular offsets, the mean delay increases, particularly at 38 GHz, indicating a greater contribution from delayed signal components arising from scattering by the complex bush structure [7], [8].

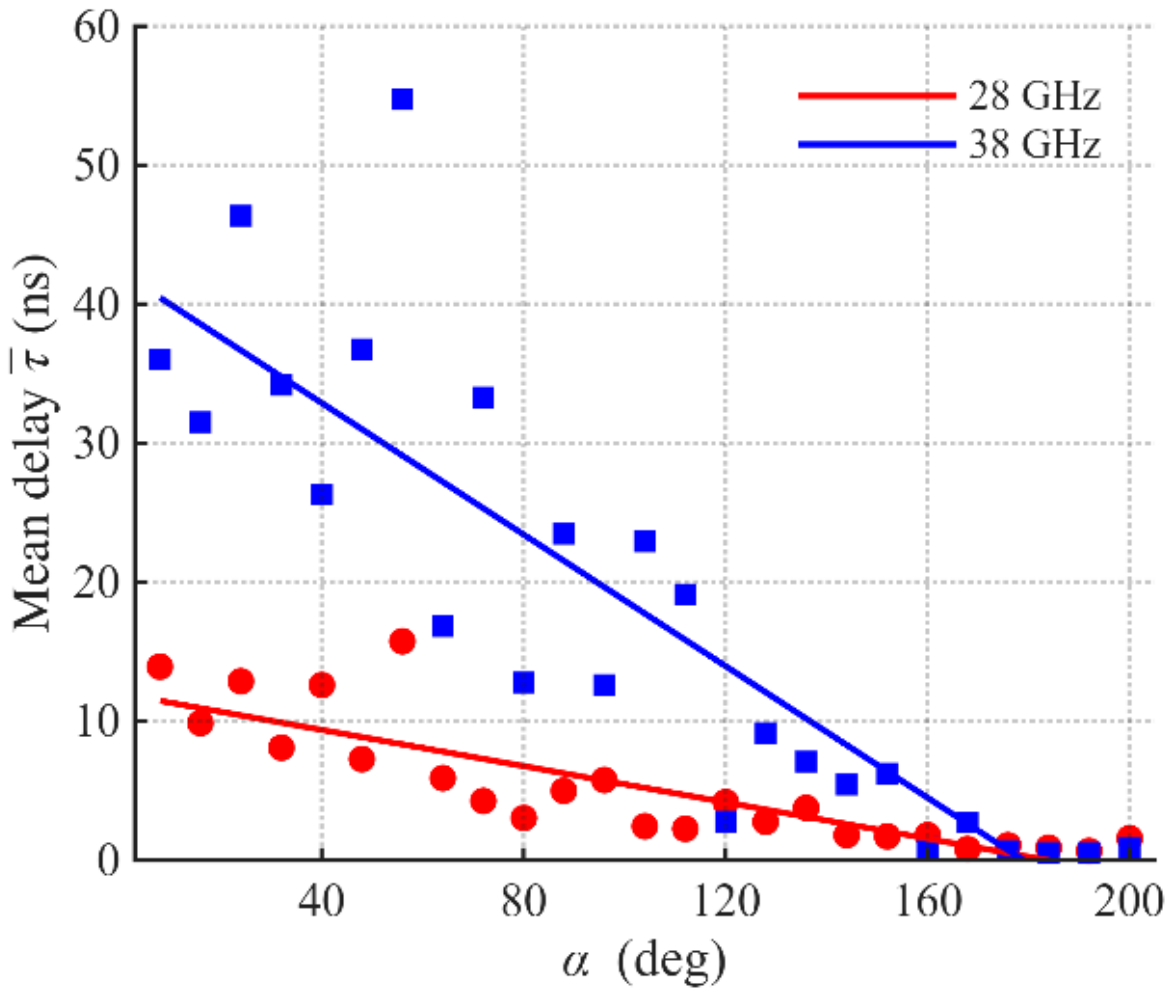

Fig. 6. Mean delay versus TX-bush-RX angle obtained from outdoor measurements at 28 GHz and 38 GHz.

## B. RMS Delay Spread

The corresponding RMS delay spread results are presented in Fig. 7. Low RMS delay spread values are observed near the main propagation direction of the TX antenna, whereas a pronounced increase in temporal dispersion occurs for off-axis angles. This effect is more pronounced at 38 GHz, highlighting frequency-dependent scattering mechanisms that remain observable even under leaf-off winter conditions [6], [9].

The angular dependence of the RMS delay spread is modeled using a first-order linear approximation given by

$$\sigma_\tau(\alpha) = F\alpha + B, \; 0° \leq \alpha \leq 180° \quad (3)$$

where $\sigma_\tau(\alpha)$ denotes the RMS delay spread expressed in nanoseconds, $\alpha$ is the antenna rotation angle, $F$ and $B$ are the model parameters obtained via linear regression, where $F$ is a slope coefficient representing the angular sensitivity of the multipath delay dispersion, and $B$ as the intercept point corresponding to the RMS delay spread at $\alpha = 0°$.

The model was estimated separately for measurements carried out at 28 GHz and 38 GHz. Table I contains the regression parameters obtained by the least-squares method.

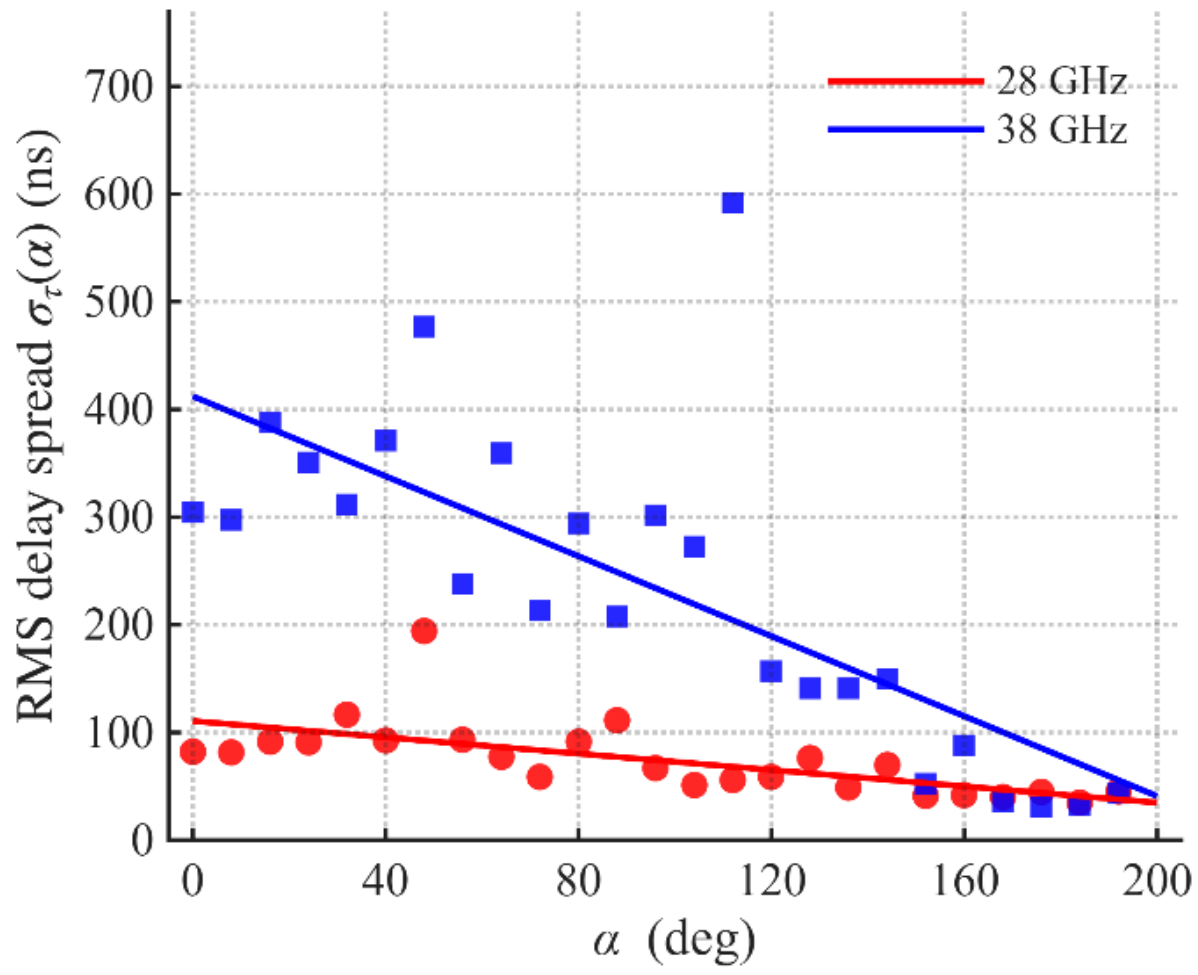

Fig. 7. RMS delay spread versus TX-bush-RX angle obtained from outdoor measurements at 28 GHz and 38 GHz.

TABLE I. Vegetation-Induced Angular RMS Delay Spread Model Parameters.

| Frequency (GHz) | Model parameters | |
|---|---|---|
| | $F$ (ns/deg) | $B$ (ns) |
| 28 | 0.067 | 110.580 |
| 38 | 0.228 | 412.200 |

The obtained linear models indicate that the RMS delay spread decreases with increasing antenna rotation angle, with a significantly higher angular sensitivity observed at 38 GHz compared to 28 GHz, reflecting stronger angular selectivity of the multipath propagation at higher frequencies.

## C. Path Loss

Fig. 8 presents the PL as a function of the observation angle at 28 GHz and 38 GHz. Minimum PLs are observed near $\alpha = 180°$, corresponding to the antenna boresight alignment, where the measured PL approaches the free-space PLs (FSPLs) of 87.4 dB and 90.1 dB for 28 GHz and 38 GHz, respectively, assuming a 20 m link distance.

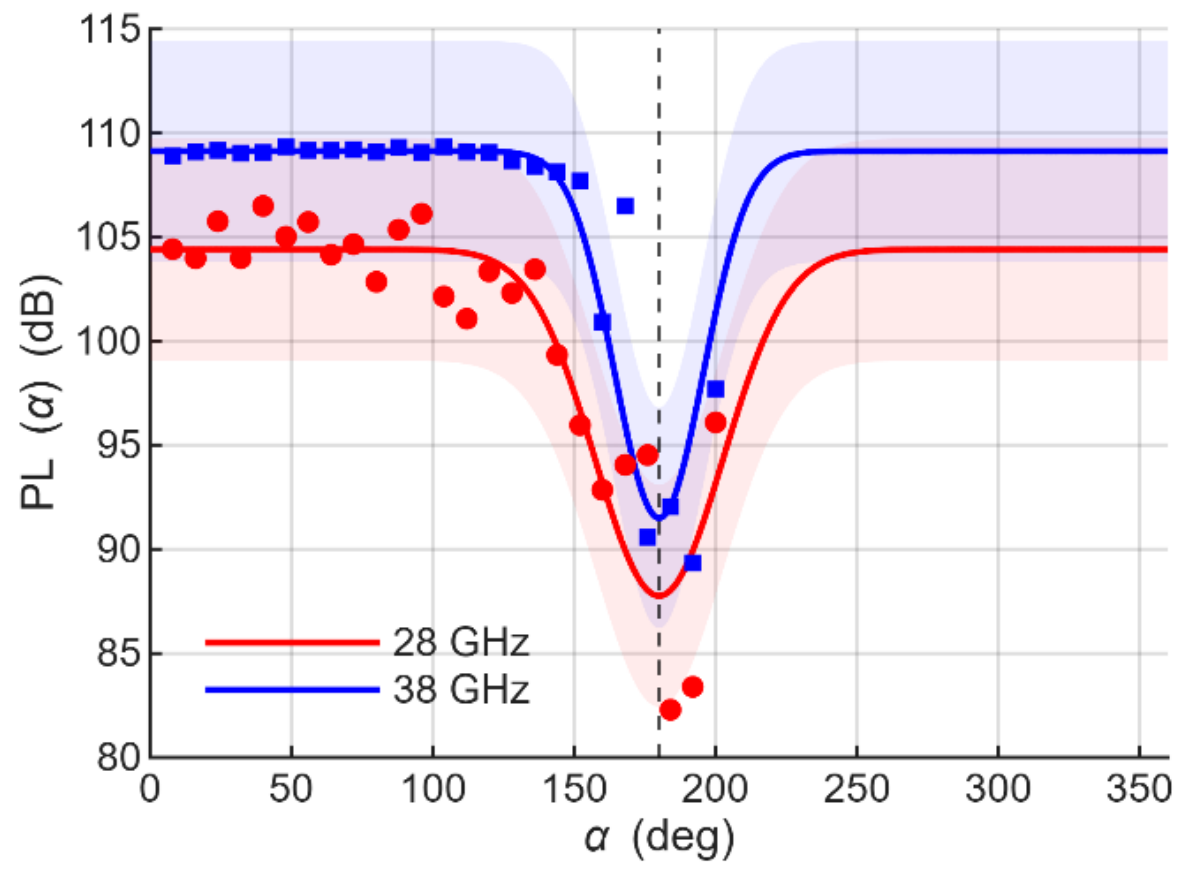

Fig. 8. PL versus TX-bush-RX angle obtained from outdoor measurements at 28 GHz and 38 GHz.

The proposed model describes the angular dependence of the PL in the presence of a single vegetation object and is given by

$$PL(\alpha) = C - A\,exp\left(-\left(\frac{\alpha-\pi}{\sigma}\right)^2\right),\ 0 \le \alpha \le \pi \quad (4)$$

where $C$ denotes the background PL level, $A$ represents the maximum PL reduction due to direct transmission through the bush, and $\sigma$ defines the angular width of the transition region around the boresight direction ($\alpha = \pi$). In this case, the angular variable $\alpha$ is expressed in radians and varies from 0 to $2\pi$. The model describes a symmetric angular transition from QLOS to scattering-dominated propagation.

Table II presents the model parameters adjusted using measurement data for individual frequencies.

TABLE II. VEGETATION-INDUCED ANGULAR PL MODEL PARAMETERS.

| **Frequency** (GHz) | **Model parameters** | | |
|---|---|---|---|
| | $C$ (dB) | $A$ (dB) | $\sigma$ (rad) |
| 28 | 104.40 | 16.65 | 0.56 |
| 38 | 109.13 | 17.61 | 0.39 |

The solid curves represent the fitted angular-dependent PL model, while the shaded regions illustrate the measurement spread defined as a constant interval of $\pm 2$ standard deviations around the model prediction. The width of the shaded regions is independent of the observation angle. For both investigated frequencies (28 GHz and 38 GHz), the standard deviation of the measurements around the mean value is approximately 2.6 dB. Assuming a normal distribution of the measurement data, the $\pm 2$ standard deviation bounds encompass approximately 95% of the measured values, which is consistent with the observed data distribution.

The difference between the FSPL and the fitted model reflects the additional attenuation introduced by the vegetation object. The higher overall attenuation observed at 38 GHz highlights frequency-dependent propagation effects resulting from the interaction between the millimeter-wave signal and the bush structure [8], [9].

## V. CONCLUSION

This paper presented an experimental analysis of angular-dependent attenuation and temporal dispersion in mmWave propagation in the presence of a single deciduous bush under winter, leaf-off conditions. Measurements conducted at 28 GHz and 38 GHz demonstrated a strong dependence of PL, mean excess delay, and RMS delay spread on the relative angular orientation of the transmitting and receiving antennas.

Minimum delay metrics and reduced temporal dispersion were observed in the main propagation direction. As the angular offset increased, both attenuation and delay-domain parameters exhibited a clear growth, indicating an increasing contribution of scattered and diffracted signal components. These effects were more pronounced at 38 GHz, confirming the frequency-dependent nature of vegetation-induced propagation mechanisms even in the absence of foliage [4], [7], [8], [9].

The results show that even isolated vegetation objects can significantly influence the angular and temporal characteristics of FR2 radio links. By incorporating angular behavior and delay-domain effects, the presented findings complement existing vegetation-attenuation models and provide valuable insights for the design and modeling of directional mmWave systems operating in environments with sparse vegetation [4], [5], [6].

## REFERENCES

[1] N. Chukhno *et al.*, "Models, methods, and solutions for multicasting in 5G/6G mmWave and sub-THz systems," *IEEE Communications Surveys & Tutorials*, vol. 26, no. 1, pp. 119–159, 2024, doi: 10.1109/COMST.2023.3319354.

[2] A. E. C. Redondi, C. Innamorati, S. Gallucci, S. Fiocchi, and F. Matera, "A survey on future millimeter-wave communication applications," *IEEE Access*, vol. 12, pp. 133165–133182, 2024, doi: 10.1109/ACCESS.2024.3438625.

[3] "Recommendation ITU-R P.833-10: Attenuation in vegetation," International Telecommunication Union (ITU), Geneva, Switzerland, Rec. ITU-R P.833-9, Sep. 2021.

[4] G. L. Ramos *et al.*, "Polarimetric vegetation propagation measurements on a Brazilian university campus scenario," *IEEE Access*, vol. 12, pp. 181498–181508, 2024, doi: 10.1109/ACCESS.2024.3507535.

[5] K. Cichoń, M. Nikiforuk, and A. Kliks, "Vegetation loss measurements for single alley trees in millimeter-wave bands," *Sensors*, vol. 24, no. 10, p. 3190, Jan. 2024, doi: 10.3390/s24103190.

[6] C. Lai, D. Senic, C. Gentile, J. Senic, and N. Golmie, "Raytracing digital foliage at millimeter-wave: A case study on calibration against 60-GHz channel measurements on summer and winter trees," *IEEE Access*, vol. 11, pp. 145931–145943, 2023, doi: 10.1109/ACCESS.2023.3345248.

[7] P. Zhang, B. Yang, C. Yi, H. Wang, and X. You, "Measurement-based 5G millimeter-wave propagation characterization in vegetated suburban macrocell environments," *IEEE Transactions on Antennas and Propagation*, vol. 68, no. 7, pp. 5556–5567, Jul. 2020, doi: 10.1109/TAP.2020.2975365.

[8] J. Cui *et al.*, "Measurement and modeling on terahertz channel propagation through vegetation," *IEEE Transactions on Terahertz Science and Technology*, vol. 15, no. 4, pp. 582–595, Jul. 2025, doi: 10.1109/TTHZ.2025.3570051.

[9] J. Wojtuń *et al.*, "Variability of radio signal attenuation by single deciduous tree versus reception angle at 80 GHz," in *2025 19th European Conference on Antennas and Propagation (EuCAP)*, Stockholm, Sweden, Mar. 2025, pp. 1–5. doi: 10.23919/EuCAP63536.2025.10999896.

[10] R. Zavorka *et al.*, "Characterizing the 80 GHz channel in static scenarios: diffuse reflection, scattering, and transmission through trees under varying weather conditions," *IEEE Access*, vol. 12, pp. 144738–144749, 2024, doi: 10.1109/ACCESS.2024.3472003.

[11] J. Blumenstein *et al.*, "Stochastic 3-D foliage modeling at 80 GHz: Experimental validation and ray-tracing simulations," *IEEE Antennas and Wireless Propagation Letters*, vol. 24, no. 10, pp. 3669–3673, Oct. 2025, doi: 10.1109/LAWP.2025.3600373.

[12] Eravant, "SAR-2507-28-S2. WR-28 Pyramidal horn antenna, 25 dBi gain," Eravant, Torrance, CA, USA, datasheet rev. 1.3, 2023. Accessed: Feb. 17, 2026. [Online]. Available: https://sftp.eravant.com/content/datasheets/SAR-2507-28-S2.pdf

[13] Eravant/SAGE Millimeter, Inc., "SBP-2034434526-2F2F-S1. Coaxial amplifier, 20 to 44 GHz, 45 dB gain, +26 dBm P1dB," Eravant, Torrance, CA, USA, datasheet rev. 1.1, 2020. Accessed: Feb. 17, 2026. [Online]. Available: https://sftp.eravant.com/content/datasheets/SBP-2034434526-2F2F-S1.pdf

[14] J. Wojtuń *et al.*, "How temporal dispersion of radio channels depends on signal bandwidth," in *2026 20th European Conference on Antennas and Propagation (EuCAP)*, Dublin, Ireland, Apr. 2026, pp. 1–5.